\documentclass[aps,pra,twocolumn,english,preprint,10pt,nofootinbib]{revtex4-2}
\usepackage[T1]{fontenc}
\usepackage[latin9]{inputenc}
\usepackage{babel}
\usepackage{nccmath}

\usepackage{setspace}
\usepackage{booktabs}
\usepackage{graphicx,amsmath,amssymb,color}
\usepackage[normalem]{ulem}
\usepackage{amsfonts}
\usepackage[toc,page]{appendix}
\usepackage{hyperref}
\usepackage{latexsym}
\usepackage{amsfonts}
\usepackage{algpseudocode}
\usepackage{booktabs}
\usepackage{amsthm}
\usepackage{mathrsfs}
\usepackage{array,multirow,tabu}
\usepackage{natbib}
\usepackage{color,verbatim}
\usepackage{psfrag}
\usepackage{orcidlink}

\usepackage{xcolor}

\def\be{\begin{equation}}
	\def\ee{\end{equation}}
\def\bea{\begin{eqnarray}}
	\def\eea{\end{eqnarray}}

\preprint{ }

\begin{document}
\setstretch{1.15}

\title{Ultra-light dark matter explanation of NANOGrav observations}

\author{Debtosh Chowdhury\orcidlink{0000-0002-4302-7356}}
\email{debtoshc@iitk.ac.in}
\address{Department of Physics, Indian Institute of Technology Kanpur, Kanpur 208016, INDIA}

\author{Arpan Hait\orcidlink{0000-0001-7540-0111}}
\email{arpan20@iitk.ac.in}
\address{Department of Physics, Indian Institute of Technology Kanpur, Kanpur 208016, INDIA}

\author{Subhendra Mohanty\orcidlink{0000-0003-0070-6647}}
\email{mohantys@iitk.ac.in}
\address{Department of Physics, Indian Institute of Technology Kanpur, Kanpur 208016, INDIA}

\author{Suraj Prakash\orcidlink{0000-0002-0917-3835}}
\email{surajprk@iitk.ac.in}
\address{Department of Physics, Indian Institute of Technology Kanpur, Kanpur 208016, INDIA}
\address{Departament de Física Teòrica, IFIC (Universitat de València - CSIC), Parc Científic UV, C/ Catedrático José Beltrán 2, E-46980 Paterna (Valencia), Spain}

\renewcommand{\abstractname}{\vspace{0.1cm}}

\begin{abstract}
	The angular correlation of pulsar residuals observed by NANOGrav~\cite{NANOGrav:2023gor} and other pulsar timing array (PTA) collaborations show evidence in support of the Hellings-Downs correlation~\cite{HD} expected from stochastic gravitational wave background (SGWB). In this paper, we offer a non-gravitational wave explanation of the observed pulsar timing correlations as caused by an ultra-light $L_{\mu} - L_{\tau}$ gauge boson dark matter (ULDM). ULDM can affect the pulsar correlations in two ways. The gravitational potential of vector ULDM gives rise to a Shapiro time delay of the pulsar signals and a non-trivial angular correlation (as compared to the scalar ULDM case). In addition, if the pulsars have a non-zero charge of the dark matter gauge group, then the electric field of the local dark matter causes an oscillation of the pulsar and a corresponding Doppler shift of the pulsar signal. We point out that pulsars carry a significant charge of muons, and thus the $L_{\mu} - L_{\tau}$ vector dark matter contributes to both the Doppler oscillations and the time delay of the pulsar signals. The synergy between these two effects provides a better fit to the shape of the angular correlation function, as observed by the NANOGrav collaboration, compared to the standard SGWB explanation or the SGWB combined with time delay explanations. Our analysis shows that in addition to the SGWB signal, there may potentially be excess timing residuals attributable to the $L_{\mu} - L_{\tau}$ ULDM.
 \end{abstract}

\maketitle

\section{Introduction} 
Precision timing of the signals from the array of pulsars with observations of over a decade provides a method for the detection of stochastic gravitational waves in the nanohertz frequency range. In their analysis of pulsar timing residuals, the NANOGrav~\cite{NANOGrav:2023gor}, EPTA~\cite{EPTA:2023fyk}, InPTA~\cite{Tarafdar:2022toa}, PPTA~\cite{Reardon:2023gzh} and CPTA~\cite{Xu:2023wog} experimental collaborations have observed a common red noise among the pulsars whose angular correlation agrees with the Hellings-Downs curve~\cite{HD} expected from an isotropic stochastic background of gravitational waves. From the 15-year pulsar-timing data set correlated among 67 pulsars, the NANOGrav collaboration~\cite{NANOGrav:2023gor} finds evidence for a stochastic signal at $(3.5-4.0)\sigma$ statistical significance. The amplitude and spectral index of the observed stochastic gravitational wave signal can be tested with theoretical templates of predictions from inflation, phase transitions, or cosmic defects ~\cite{NANOGrav:2023hvm, Lazarides:2023ksx, Lazarides:2023rqf, Maji:2023fhv, Kitajima:2023cek}. The observations of the NANOGrav collaboration also agree with the prediction from the cumulative stochastic gravitational wave background (SGWB) signals from the merger of super-massive black holes ($\sim 10^6 M_\odot$) over cosmological times~\cite{NANOGrav:2023hfp}. In theories of gravity beyond Einstein's GR, the scalar and vector modes of the metric can propagate, and these modes will give angular correlations which deviate from the Hellings-Downs shape~\cite{non-Einstein,Cornish:2017oic}, and the NANOGrav data~\cite{NANOGrav:2023gor} have been used for constraining the non-Einsteinian modes of the metric perturbations~\cite{Chen:2023uiz,NANOGrav:2023ygs}.

The NANOGrav observations can also have explanations without gravitational waves. Ultra-light vector dark matter (ULDM), with mass $\sim 10^{-23} {\rm eV}$, induces displacements of the Earth as well as the pulsar that cause a periodic signal in timing residuals with frequency $f = m/2  \pi \simeq 2.4 \times 10^{-9}{\rm Hz}$. This frequency falls into the sensitivity region of PTA experiments with an observation time of decades~\cite{Nomura:2019cvc,PPTA:2021uzb}. A single-field dark matter model exclusively accounts for a monochromatic gravitational wave signal. The stochastic signal reported by NANOGrav~\cite{NANOGrav:2023gor}, encompassing a range of frequency, can be realized in multi-field scenarios such as the Kaluza-Klein theory with a tower of masses~\cite{Anchordoqui:2023tln} or continuous mass `unparticles'~\cite{Deshpande:2008ra}. Examples of such ultralight dark photons which can couple to the Earth and/or pulsars are gauge bosons arising from $U(1)_{B}$ or $U(1)_{B-L}$ symmetries~\cite{Marshak:1979fm, Fayet:1989mq,  Fayet:1990wx, Khalil:2006yi, Basso:2008iv, Iso:2009ss, Okada:2010wd, Basso:2010jm, Basso:2010yz, Kanemura:2011vm, Fayet:2017pdp, Fayet:2018cjy}. 

The local Newtonian potential of Ultra-light dark matter can also cause a gravitational frequency shift in the pulsar signals, which affects the angular correlations and whose shape depends on the spin of the ULDM~\cite{Sun:2021yra,Unal:2022ooa,PPTA:2022eul,Xia:2023hov,Omiya:2023bio}. The gravitational time delay from ULDM will occur even if the DM particles have no coupling to the Earth or Pulsar. 

In this paper, we study the $L_\mu-L_\tau$ ULDM. In this case, both the aforementioned effects are present. Neutron stars carry substantial muon charges and are therefore oscillated by the background dark matter~\cite{Pearson:2018tkr, Potekhin:2013qqa, Garani:2019fpa}. The oscillation of the Earth term is absent as the Earth is expected to have zero muon charge. In addition to oscillating the pulsars, the $L_\mu-L_\tau$ ULDM also gives rise to time delay due to gravitational potential perturbation. We find that taking both the time delay and the source oscillations together provides a better fit to the shape of the angular correlation function, as found by the NANOGrav collaboration, compared to the Hellings-Downs curve from SGWB. 

The  $L_\mu-L_\tau$ gauge theory is of interest as it is an anomaly-free extension of the standard model without the need of adding extra families of fermions for anomaly cancellation~\cite{Foot:1990mn,He:1991qd,Foot:1994vd, Heeck:2011wj}. The $L_\mu-L_\tau$ interaction is described by the Lagrangian
\begin{eqnarray}
	{\cal L}=- g^\prime Z^\prime_\alpha\left( \bar \mu \gamma^\alpha \mu -\bar \tau \gamma^\alpha \tau + \bar\nu_\mu\gamma^\alpha \nu_\mu - \bar\nu_\tau\gamma^\alpha \nu_\tau\right).
\end{eqnarray} 

These interactions can give rise to long-range forces between leptons, which may be probed via neutrino oscillations~\cite{Grifols:2003gy,Joshipura:2003jh,Heeck:2010pg,Farzan:2016wym,Smirnov:2019cae,Dror:2020fbh,Singh:2023nek} or binary pulsar timings~\cite{KumarPoddar:2019ceq,Kopp:2018jom}. The $L_\mu-L_\tau$ gauge bosons, if they are ultra-light, can also serve as the fuzzy dark matter of the universe~\cite{ Hu:2000ke,Hui:2016ltb,Brzeminski:2022rkf}.

Since planets and stars do not carry any $L_\mu-L_\tau$ charge, the long-range gauged forces associated with this charge cannot be probed from planetary or stellar dynamics, unlike other anomaly-free combinations of charges like $L_e-L_{\tau/\mu}$ or $B-L$ which are sourced by the electron or baryon content of these bodies~\cite{KumarPoddar:2020kdz}. However, neutron stars carry a substantial number of muons, which can be a source of ultra-light gauge boson radiation from binary neutron stars, and these can be radiated in binary stars. Therefore, their mass and couplings can be constrained from binary pulsar timings~\cite{KumarPoddar:2019ceq,Kopp:2018jom}. The best-fit values deduced through our analysis of the mass and couplings of the $L_\mu-L_\tau$ vector field fall within the allowed parameter space based on the measured loss in the time periods of such binary pulsars.

We introduce the vector dark matter model and evaluate the correlations of pulsar timing residuals in section \ref{sec:DM-and-residual}. We then conduct a detailed comparison of the angular correlations due to different sources, i.e., the stochastic gravitational wave background, the gravitational potential of the dark matter, and from pulsar oscillations due to their coupling with the dark matter in section \ref{sec:angular-corr-comp-fit}. We also identify the best fit to data among the different theoretical models through a $\chi^2$ minimization. We then translate the results of the fitting with data into constraints on the dark matter mass and coupling parameters in section \ref{sec:param-constraints}. Finally, we summarize the analysis and present our conclusions in section \ref{sec:conclusions}.

\section{Pulsar period residual from Proca dark matter}\label{sec:DM-and-residual}

We consider the case when neutron stars carry non-zero charges of the gauge group of the Proca field. This is true if the gauge group is $B$ or $B-L$. This is also true for gauge groups like $L_\mu-L_\tau$ as neutron stars carry a large number of muons whose decay is prevented by Fermi-blocking of the outgoing electrons due to electron degeneracy~\cite{KumarPoddar:2019ceq,Kopp:2018jom}. 

\subsection{Proca dark matter}

The Lagrangian for a massive vector field is given by
\begin{eqnarray}
	{\cal L}= -\cfrac{1}{4}\,Z^{\prime}_{\mu \nu}\,Z^{\prime\mu \nu} +\cfrac{1}{2}\,m^2 Z^{\prime}_\mu \, Z^{\prime \mu} + g\, J_\mu \, Z^{\prime \mu}\,.
\end{eqnarray}

From this we obtain the equation of motion for the vector field $Z^{\prime}_\mu(\vec x,t)$ in free space ($J^\mu=0$) given by
\begin{eqnarray}
	\partial_\mu\, Z^{\prime \mu \nu} +m^2 Z^{\prime \nu}=0,
\end{eqnarray}

which reduces to the following Eqs.,
\begin{eqnarray}
	&&\left(\partial_t^2 -\nabla^2 +m^2 \right)Z^{\prime \mu}=0\,,\\
	&& \partial_t\, Z^{\prime t} -\nabla\cdot \vec Z^{\prime}=0\, \label{eom2}.
\end{eqnarray}

From Eq.~\eqref{eom2}, we see that $Z^{\prime t}= (k_i/m) Z^{\prime i}$. If the massive vector boson constitutes dark matter then the momenta $k_i \sim m v_i$, where the dark matter velocity in the galaxy $v^i\sim 10^{-3}$. This implies that $Z^{\prime t} = v_i Z^{\prime i}$. Thus the spatial components dominate $Z^{\prime}_{i} \gg Z^{\prime}_t$ and we can take $Z^{\prime}_\mu\simeq (0, \vec Z^{\prime})$ for the dark matter vector field. The energy density is given by the time component of the stress tensor,
\begin{eqnarray}
	T_{00}\simeq  \cfrac{1}{2}\, m^2\, \vert \vec Z^{\prime}\vert^2= \rho_\text{dm},
\end{eqnarray}
where the dark matter density in the galaxy is $\rho_\text{dm}= 0.4 \,{\rm GeV/cm^3}$. The vector field for the Proca dark matter can be written as~\cite{Nomura:2019cvc}
\begin{eqnarray}
	\vec Z^{\prime}(\vec x,t)= \cfrac{\sqrt{2 \rho_\text{dm}}}{m}\, \cos (mt+\vec k\cdot \vec x),
\end{eqnarray}
with $|\vec k|= m\,v$.

\subsection{Angular correlations of pulsar timing residuals}

A neutron star of mass $M_a$ carrying a gauge charge $Q_a$ of the dark matter will experience a time-dependent force due to the dark `electric field' given by
\begin{eqnarray}\label{force}
	\vec F=\sum_{A=1,2}  Q_a\, m\, \vec \epsilon_A \left(\vec n\right)\,\cfrac{\sqrt{2 \rho_{\rm dm}}}{m}\, \sin\left(mt+\vec k\cdot \vec x\right),
\end{eqnarray}
where $\vec \epsilon_A \left(\vec n\right)$ is the polarization vector of the gauge field. We have used the fact that the vector field for the dark matter has the form
\begin{eqnarray}
	\vec A\left(\vec x,t\right)= \cfrac{\sqrt{2 \rho_{\rm dm}}}{m}\, \cos \left(mt+\vec k\cdot \vec x\right),
\end{eqnarray}
with $|\vec k|= m\,v$. In time $\delta t$, the neutron star has a perturbation in its position given by
\begin{eqnarray}
	\delta \vec x= - \sum_{A=1,2}\, \vec \epsilon_A \left(\vec n\right) \, \cfrac{Q_a}{M_a\, m}\, \cfrac{\sqrt{2 \rho_{\rm dm}}}{m} \, \sin \left(mt+\vec k\cdot \vec x \right).
\end{eqnarray}
For a pulsar located in the direction $\hat n_a$ and at a distance $c\,t_a$ from the Earth having time period $T_a$, the shift in the period due to the perturbation is given as
\begin{eqnarray}
	\Delta T_a \left(t\right)&=& \hat n_a \cdot \left[\,\delta \vec x\left(t-t_a+T_a\right)-\delta \vec x\left(t-t_a\right)\,\right]\\
	&\simeq&  \sum_{A=1,2} \, T_a \,  \hat n_a \cdot  \vec \epsilon_A \left(\vec n\right)\, \cfrac{Q_a}{M_a} \, \cfrac{\sqrt{2 \rho_{\rm dm}}}{m} \, \cos \left[m\left(t-t_a\right)\right]. \nonumber
\end{eqnarray}
Defining the pulsar period redshift and the pulsar timing residual as
\begin{eqnarray}
	z_a(t)\,\equiv\, -\cfrac{\Delta T_a}{T_a}, \quad \text{and} \quad R_a(t) = \int_0^t\,z_a(t^\prime)\,dt^\prime.
\end{eqnarray}
We can compute the two point correlation of period residuals of two pulsars with angular separation $\theta_{ab}$ as
	\begin{eqnarray}
		\langle R_a(t) R_b(t+\tau) \rangle= \cfrac{Q_a Q_b}{M_a M_b}\, \cfrac{ \rho_{\rm dm}}{m^4}\, \cfrac{2}{3} \cos\theta_{ab}\cos(m\tau).
		\label{deltaT-ab}
	\end{eqnarray}
	\noindent Here the angular brackets denote averaging over time and the directions of the DM Proca field propagation direction $\vec n$. We have made use of the relation 
	\begin{eqnarray}\label{eq:integ-angular}
		\int\, \cfrac{d^2 \vec n}{4 \pi}\, \sum_A\, \epsilon_A^i (\vec n)\, \epsilon_A^j(\vec n)\, n^i_a \, n^j_b = \cfrac{2}{3} \, \cos \theta_{ab},
	\end{eqnarray}

which we have derived in appendix~\ref{app:angular-integ}
	
	\section{Stochastic Gravitational Waves vs Dark Matter}\label{sec:angular-corr-comp-fit}
	
	\subsection{Angular correlations due to distinct sources}
	
	We compare three distinct theoretical models aimed at explaining the experimentally reported angular correlations:
	
	{\bf Case 1:} When the stochastic gravitational wave background (SGWB) is the only source of pulsar oscillations, the angular correlation of the timing residual can be written as a harmonic sum of the SGWB frequency$f_i$~\cite{NANOGrav:2023gor}: 
	\begin{eqnarray}
	\xi^{(1)}_{ab}\left(\theta_{ab}\right) = \Phi_{\rm GW} \left(f_i\right)\Gamma_{\rm HD}\left(\theta_{ab}\right), 
	\end{eqnarray}
	where the amplitude due to stochastic gravitational waves can be written as~\cite{NANOGrav:2023gor}
	\begin{eqnarray}\label{eq:amplitude-SGW}
	\Phi_{\rm GW}\left(f_i\right)= \frac{A_{\rm GW}^2}{12 \pi^2} \frac{1}{T_{\rm obs}} \left(\frac{f_i}{f_{\rm ref}}\right)^{-\gamma} f_{\rm ref}^{-3}\,.
	\end{eqnarray}
	Here, $A_{\rm GW}$ is the average amplitude of the SGWB and $\gamma$ is the spectral index which depends on the nature of the source and for binary black-hole mergers $\gamma=13/3$~\cite{Phinney:2001di}.
	
	The angular correlation for SGWB is given by the Hellings-Downs function~\cite{HD}
		\begin{eqnarray}\label{eq:HD-correlation}
			\Gamma_{\rm HD}\left(\theta_{ab}\right) &=& \cfrac{1}{2} + \cfrac{3}{2}\,\cfrac{\left(1-\cos\theta_{ab}\right)}{2}\, \ln  \cfrac{\left(1-\cos\theta_{ab}\right)}{2}  \nonumber\\
			&& \qquad- \cfrac{1}{4}  \cfrac{\left(1-\cos\theta_{ab}\right)}{2}.
		\end{eqnarray}
		
	{\bf Case 2:} If in addition to the pulsar oscillations due to the SGWB, time delay due to the gravitational potential of vector dark matter (the Shapiro time delay) is also taken into account, then the total angular correlation at the frequency $f_{i}= m/\pi$ can be obtained as \cite{Omiya:2023bio,Nomura:2019cvc}:
	\begin{eqnarray}\label{eq:HD+shapiro-corr-1}
		\xi^{(2)}_{ab}\left(\theta_{ab}\right) = \Phi_{\rm VDM}^{\rm GP}\, \Gamma_{\rm  VDM}^{\rm GP}\left(\theta_{ab}\right) + \Phi_{\rm{GW}}\left(m/\pi\right)\, \Gamma_{\rm HD}\left(\theta_{ab}\right),\nonumber \\
	\end{eqnarray}
	
	where $\Phi_{\rm VDM}^{\rm GP}$ is the amplitude corresponding to the shapiro time delay, and it has the form
\begin{eqnarray}\label{Phi-VDM-GP}
	\Phi^{\rm GP}_{\rm VDM} &=& \cfrac{69}{5\,\left(2m\right)^2}\, \cfrac{\pi^2 G^2}{36}\,\cfrac{4\,\rho_{\rm dm}^2}{m^4},
\end{eqnarray}
and 
	\begin{eqnarray}\label{eq:shapiro-correlation}
		\Gamma_{\rm VDM}^{\rm GP}\left(\theta_{ab}\right)
		&=& \cfrac{5}{138}\,P_{0}\left(\cos\theta_{ab}\right) + \cfrac{64}{138}\,P_{2}\left(\cos\theta_{ab}\right),
	\end{eqnarray}
	
	with $P_{n}\left(x\right)$ being the $n$-th Legendre polynomial.

	We define the ratio $\beta =  \cfrac{\Phi_{\rm GW}\left(m/\pi\right)}{\Phi_{\rm VDM}^{\rm GP}}$, and normalize the angular correlation so as to ensure that the normalized correlation function $\tilde{\xi}^{(2)}_{ab}\left(0\right) = \cfrac{1}{2}$ irrespective of the choice of $\beta$. Therefore,
	\begin{eqnarray}\label{eq:HD+shapiro-corr-2}
		\tilde{\xi}^{(2)}_{ab}\left(\theta_{ab}\right) 
		& = &\cfrac{\Phi_\text{VDM}^\text{GP}\, \Gamma_\text{VDM}^\text{GP} (\theta_{ab}) + \Phi_\text{GW}\, \Gamma_\text{HD}(\theta_{ab})}{\Phi_\text{VDM}^\text{GP} + \Phi_\text{GW}} \nonumber\\ 
		& =& \cfrac{1}{1+\beta} \left[ \Gamma^{
			\rm GP}_{\rm VDM}\left(\theta_{ab}\right) + \beta\, \Gamma_{\rm HD}\left(\theta_{ab}\right) \right].
	\end{eqnarray}
		
	{\bf Case 3:} As an alternative to the effect of the SGWB, if the source of pulsar oscillations is their coupling with the Proca field (see Eq.~\eqref{force}), on account of their muon content (see appendix \ref{app:muon-charge}), then also taking into account the Shapiro time delay, the total angular correlation has the form:
	\begin{eqnarray}\label{eq:shapiro+vdm-corr-1}
			\xi^{(3)}_{ab}\left(\theta_{ab}\right) = \Phi_{\rm VDM}^{\rm GP}\,\Gamma_{\rm VDM}^{\rm GP}\left(\theta_{ab}\right) + \Phi_{\rm VDM}\, \Gamma_{\rm VDM}\left(\theta_{ab}\right).\nonumber \\
	\end{eqnarray}
	Here, $\Phi_{\rm VDM}^{\rm GP}$ and $\Phi_{\rm VDM}$ are functions of the DM and pulsar parameters,
	\begin{eqnarray}\label{Phi-VDM}
			\Phi_\text{VDM} &=& \cfrac{4}{3}\, \cfrac{Q_a\, Q_b}{M_a\, M_b}\,   \cfrac{ \rho_{\rm dm}}{m^4}.			
	\end{eqnarray}
	The angular correlation due to the Doppler shift of the signal on account of pulsar oscillations in the vector dark matter (VDM) background is given in Eq.~\eqref{eq:integ-angular},
	\begin{eqnarray}
	\Gamma_{\rm VDM}\left(\theta_{ab}\right) &=& \cfrac{1}{2} \cos\theta_{ab},
	\end{eqnarray}
	where we have absorbed a factor of $4/3$ in the amplitude to follow the Hellings-Downs normalization, see Eq.~\eqref{eq:HD-correlation}, whereby we define the shape parameters such that $\Gamma\left(\theta_{ab}=0\right)=\frac{1}{2}$.	
Thus, we rewrite the normalized correlation function as
\begin{eqnarray}\label{eq:shapiro+vdm-corr-2}
\tilde{\xi}_{ab}^{(3)}\left(\theta_{ab}\right)
 & = & \cfrac{\Phi_\text{VDM}^\text{GP}\, \Gamma_\text{VDM}^\text{GP} (\theta_{ab}) + \Phi_\text{VDM}\, \Gamma_\text{VDM}(\theta_{ab})}{\Phi_\text{VDM}^\text{GP} + \Phi_\text{VDM}} \nonumber\\ 
& = & \cfrac{1}{1+\alpha} \left[\Gamma_{\rm VDM}^{\rm GP}\left(\theta_{ab}\right) + \alpha\,  \Gamma_{\rm VDM}\left(\theta_{ab}\right)\right],
\end{eqnarray}
with $\alpha$ being the ratio $\Phi_{\rm VDM}/\Phi_{\rm VDM}^{\rm GP}$. 

We ascertain the best-fit values of $\beta$ and $\alpha$ by means of $\chi^2$ minimization in the next section.

\subsection{Fit with NANOGrav data}

We have conducted separate fits with respect to the amplitude and shape of the reported signal. This is facilitated by the normalization of the angular correlations independent of the amplitude for each of the three cases.
	
The $\chi^2$ function corresponding to the angular correlations can be constructed in terms of the experimentally reported data and error $\left(\xi^{\rm exp}_{ab}\left(\theta_i\right), \sigma^{\rm exp}\left(\theta_i\right)\right)$ and the theoretical results $\xi^{\rm exp}_{ab}\left(\theta_i, \vec\delta\right)$ as
\begin{eqnarray}\label{eq:chi-sq-general}
	\chi^{2}(\vec\delta) = \sum_i \left(\cfrac{\xi^{\rm exp}_{ab}\left(\theta_i\right) - \xi^{\rm th}_{ab}\left(\theta_i, \vec\delta\right)}{\sigma^{\rm exp}\left(\theta_i\right)}\right)^2,
\end{eqnarray}
where $\theta_i$ designate the data points and $\vec\delta$ denotes the set of parameters to be fitted. We find that for cases
	
\begin{enumerate}
	\item For pulsar oscillations due to the SGWB, the Hellings-Downs function describes the angular correlations. As evident from Eq.~\eqref{eq:HD-correlation}, there are no unknown parameters, and one can simply evaluate the $\chi^2$ function as
		\begin{eqnarray}\label{eq:chi-sq-HD}
			\chi^{2} = \sum_i \left(\cfrac{\xi^{\rm exp}_{ab}\left(\theta_i\right) - \xi^{(1)}_{ab}\left(\theta_i\right)}{\sigma^{\rm exp}\left(\theta_i\right)}\right)^2 = 94.303.
		\end{eqnarray}
		
		\item
		
		\begin{figure}[!htb]
			\centering
			\includegraphics[width=8.5cm]{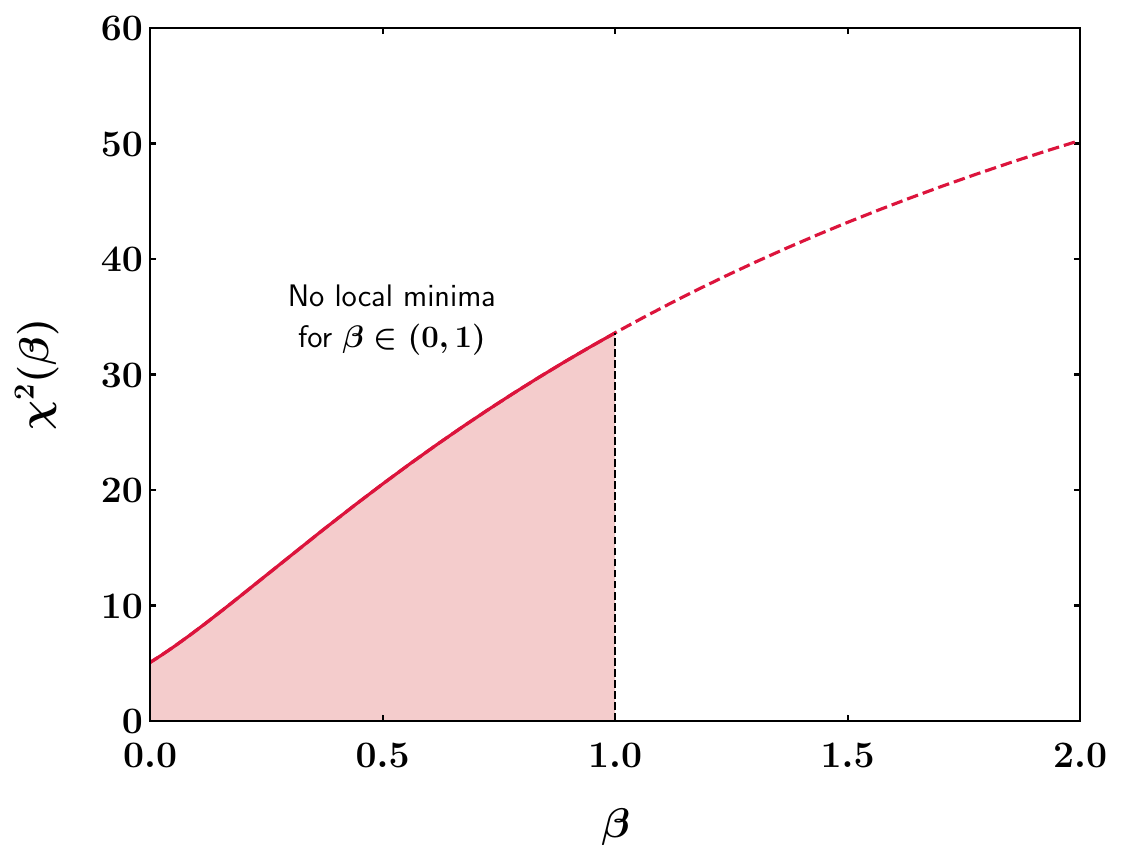}
			\caption{\textit{A plot of $\chi^{2}$ as a function of the fraction $\beta$ for the model that takes into account the combined effects of the stochastic gravitational background and the gravitational potential of vector dark matter. The highlighted region corresponds to $\beta \in (0,1)$.}}
			\label{fig:chi_sq_HD_Shapiro}
		\end{figure}
		
	For the combined effect of oscillations due to the SGWB and the Shapiro time delay owing to the gravitational potential of the VDM, the $\chi^2$ is a function of one unknown variable $\beta$, i.e.,
		\begin{eqnarray}\label{eq:chi-sq-HD+shapiro}
			\chi^{2}\left(\beta\right)
			&=& \sum_i \left(\cfrac{\xi^\text{exp}_{ab}(\theta_i) - \xi^{(2)}_{ab}(\theta_i, \beta)}{\sigma^\text{exp}(\theta_i)}\right)^2 \nonumber\\ 
			&=& \cfrac{ 5.068 + 35.052\,\beta + 94.303\, \beta^2 }{\left(1+\beta\right)^2}.
		\end{eqnarray}
		The minima is obtained for $\beta = \beta^* \rightarrow 0$ with $\chi^2_{\rm min} = 5.068$, which is a notable improvement over the $\chi^2$ corresponding to the Hellings-Downs correlation alone. The features of $\chi^2(\beta)$ have been shown in Fig.~\ref{fig:chi_sq_HD_Shapiro}.
		
		Extrapolating the $\chi^2(\beta)$ curve beyond $\beta = 1$, it can be seen that the function keeps on increasing.
		
		\item 
		
		\begin{figure}[!htb]
			\centering
			\includegraphics[width=8.5cm]{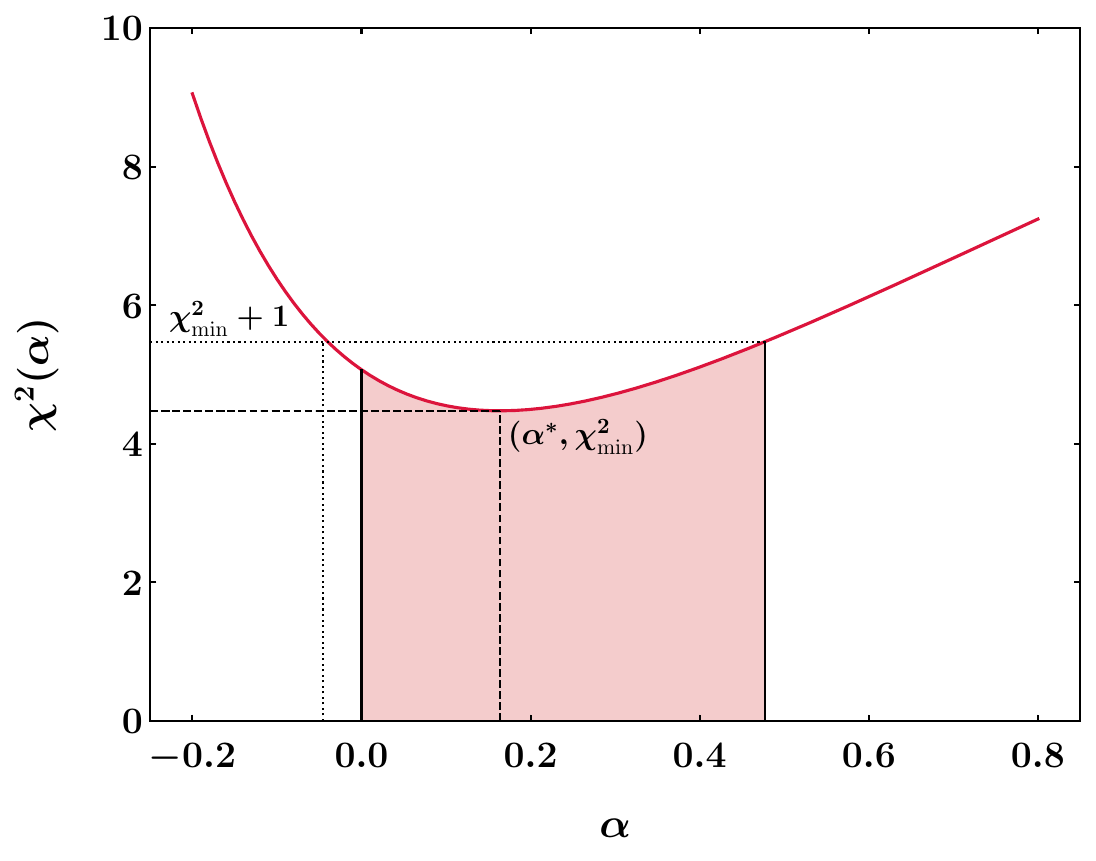}
			\caption{\textit{A plot of $\chi^{2}$ as a function of the fraction $\alpha$ for the theoretical model that takes into account the combined effects of the gravitational potential of Proca dark matter and its coupling to the pulsars. Here, $\alpha^*$ corresponds to the value that yields the minimum value of $\chi^2_\text{min}$ and the region highlighted in red corresponds to the $1\sigma$ confidence interval.}}
			\label{fig:chi_sq_f}
		\end{figure}
	
		When the source of the oscillations is the coupling of the pulsars with an $L_{\mu}-L_{\tau}$ ULDM background and if the Shapiro time delay is also taken into account, the $\chi^2$ can be expressed as a function of the variable $\alpha$, i.e.,
		\begin{eqnarray}\label{eq:chi-sq-shapiro+VDM}
			\chi^{2}\left(\alpha\right)&=& \sum_i \left(\cfrac{\xi^\text{exp}_{ab}(\theta_i) - \xi^{(3)}_{ab}(\theta_i, \alpha)}{\sigma^\text{exp}(\theta_i)}\right)^2 \nonumber\\ 
			&=& \cfrac{5.068 + 1.706\,\alpha + 26.612\, \alpha^2 }{\left(1+\alpha\right)^2}.
		\end{eqnarray}
		The minima is obtained at $\alpha = \alpha^* = 0.16365$ with $\chi^{2}_{\rm min} \approx 4.47$, which improves over both the previous cases. The features of $\chi^2(\alpha)$ along with the minima have been distinctly highlighted in Fig.~\ref{fig:chi_sq_f}, where we have also highlighted the range of $\alpha$ values that
		correspond to the $1\sigma$ confidence interval. Since negative values of $\alpha$ do not carry any physical meaning, the lower limit of the $1\sigma$ interval is truncated at $\alpha = 0$.
		
	\end{enumerate}
	
\begin{figure}[!htb]
			\centering
			\includegraphics[width=8.5cm]{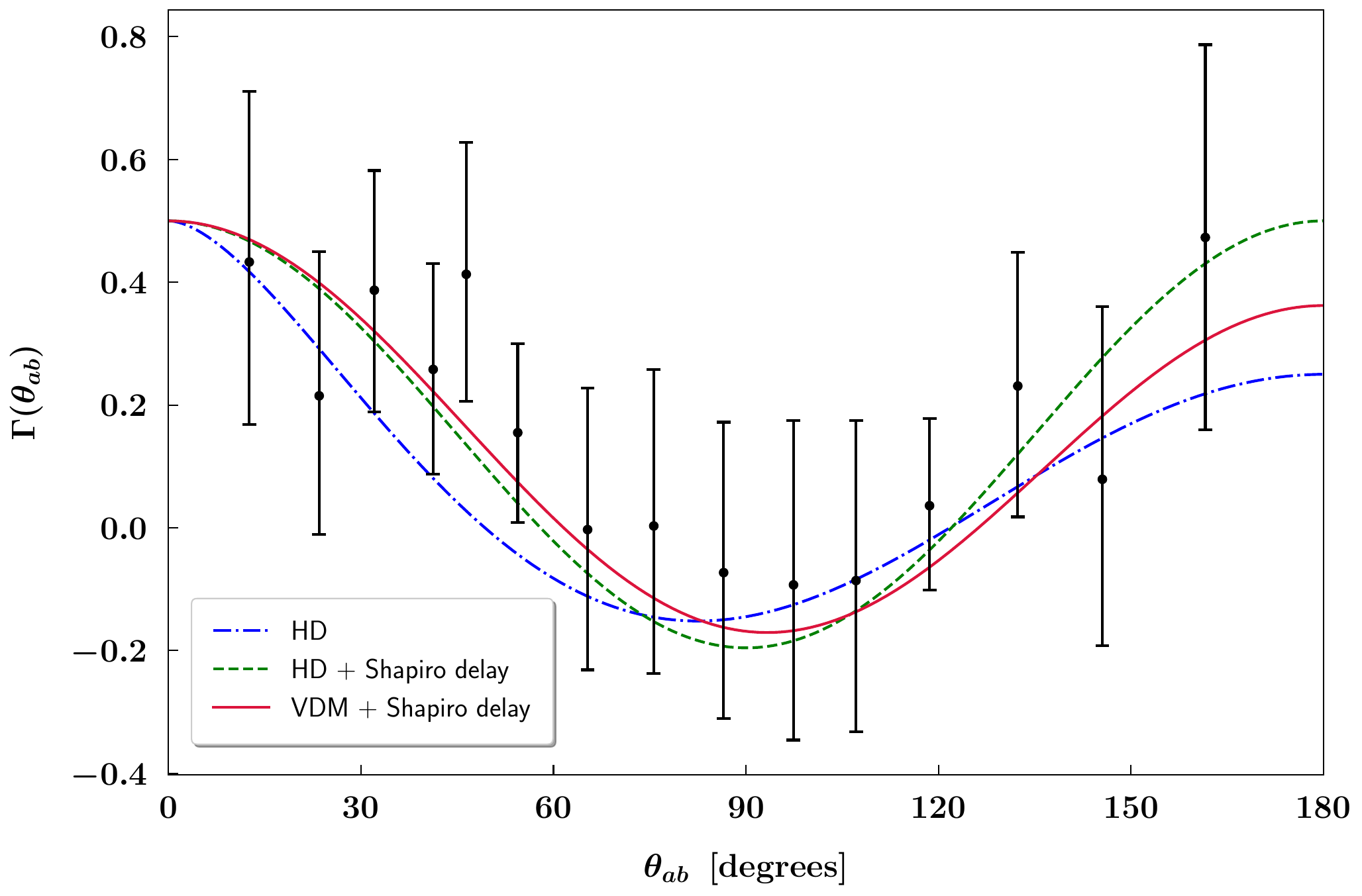}
			\caption{\textit{A comparison of the three theoretical models describing the angular correlation through: (1) The Hellings-Downs function with $\chi^2 = 94.303$, (2) The best fit curve for the scenario involving the Hellings-Downs function along with the correlation due to the time delay caused by the gravitational potential of Proca dark matter, with $\chi^2_{\rm min} = 5.068$, (3) Angular correlation due to the gravitational potential as well as the coupling of the Proca dark matter with the pulsars with $\chi^2_{\rm min} = 4.47$. We have also shown the experimentally reported data by black dotted points along with their error bars \cite{NANOGrav:2023gor}.}}
			\label{fig:angular_correlation}
		\end{figure}

	The angular correlations corresponding to these three cases have been plotted against the experimentally reported $\left(\theta_i, \xi^{\rm exp}_{ab}\left(\theta_i\right)\right)$ values~\cite{NANOGrav:2023gor} in Fig.~\ref{fig:angular_correlation}.

		\section{Constraints on the Parameters}\label{sec:param-constraints}
		
		\begin{figure*}[!htb]
			\centering
			\includegraphics[width=15cm]{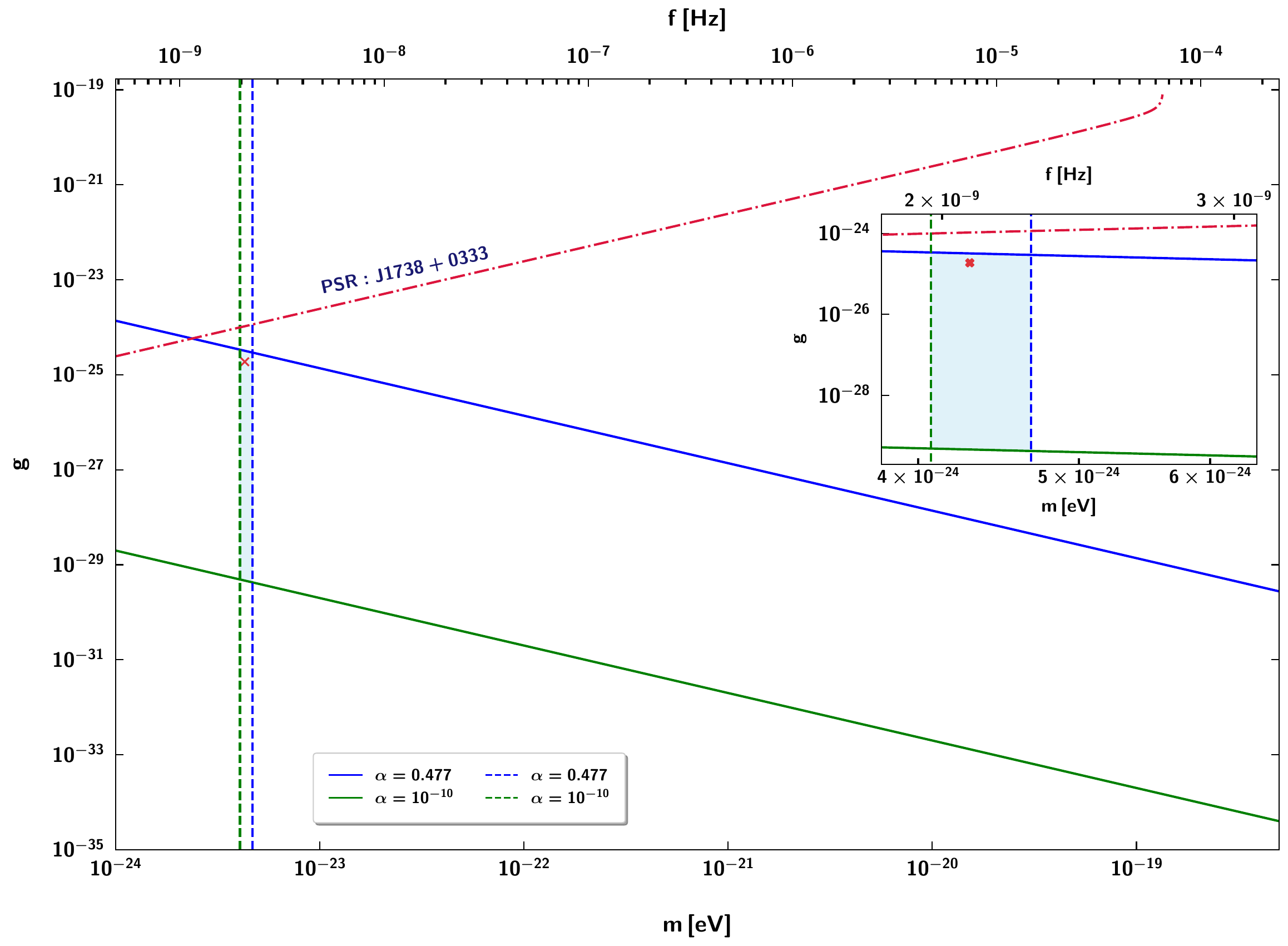}
			\caption{\it The allowed range of $g$ vs $m$ values based on constraints from the amplitude as well as the shape of the reported signal \cite{NANOGrav:2023gor}. The solid lines correspond to contours obtained from Eq.~\eqref{eq:alpha-num} after substituting $\alpha$ values corresponding to the boundaries of the $1\sigma$ region. The lower boundary is truncated at $\alpha \sim 0$ since negative values of $\alpha$ are unphysical. The vertical dashed lines highlight the constraints on the dark matter mass from the amplitude of the signal. An enlarged view of the overlap region, along with the best fit-point $(m = 4.297\times10^{-24} {\rm eV},\, g = 1.876\times10^{-25})$ corresponding to $\alpha = 0.16365$ has been shown as an inset. The $(g,m)$ parameter space above the dashed-dotted curve is ruled out based on the measurements of binary pulsar timings \cite{KumarPoddar:2019ceq}.}
			\label{fig:g-vs-m}
		\end{figure*}
	
		The ratio $\alpha$ can be related to the DM parameters, i.e., the coupling $g$ and the mass $m$ by noting that from Eqs.~\eqref{Phi-VDM-GP} and~\eqref{Phi-VDM},

		\begin{eqnarray}
			\alpha = \cfrac{\Phi_{\rm VDM}}{\Phi_{\rm VDM}^{\rm GP}} =  \cfrac{80}{23\pi^2} \,\cfrac{Q_a\,Q_b}{M_a\,M_b\,G^2}\,\cfrac{m^2}{\rho_{\rm dm}}.
		\end{eqnarray}
	After substituting the following numerical values~\cite{KumarPoddar:2019ceq}, 
	\begin{eqnarray}\label{eq:num-vals}
		&&	N = 10^{55},\quad	M_a = M_b = 10^{57}\, \text{GeV},\nonumber \\ && \rho_{\rm dm} = 3.1 \times 10^{-42}\, \text{GeV}^{4}.
	\end{eqnarray}
	We obtain
	\begin{eqnarray}\label{eq:alpha-num}
		\alpha = 2.518 \times 10^{95}\, g^2\,\left(\cfrac{m}{{\rm eV}}\right)^{2}.
	\end{eqnarray}
	Substituting the best-fit value of $\alpha$ yields a contour in the $g-m$ plane. The $1\sigma$ allowed region is defined by $\alpha \in (0, 0.477]$ and this yields upper and lower\footnote{For the lower bound, we have chosen $\alpha =  10^{-10}$, since $\alpha = 0$ implies either $g = 0$ or $m = 0$ from Eq.~\eqref{eq:alpha-num}.} bounds in the $g-m$ plane and these have been highlighted by solid lines in Fig.~\ref{fig:g-vs-m}.
		
\begin{figure}[t]
			\centering
			\includegraphics[width=8.5cm]{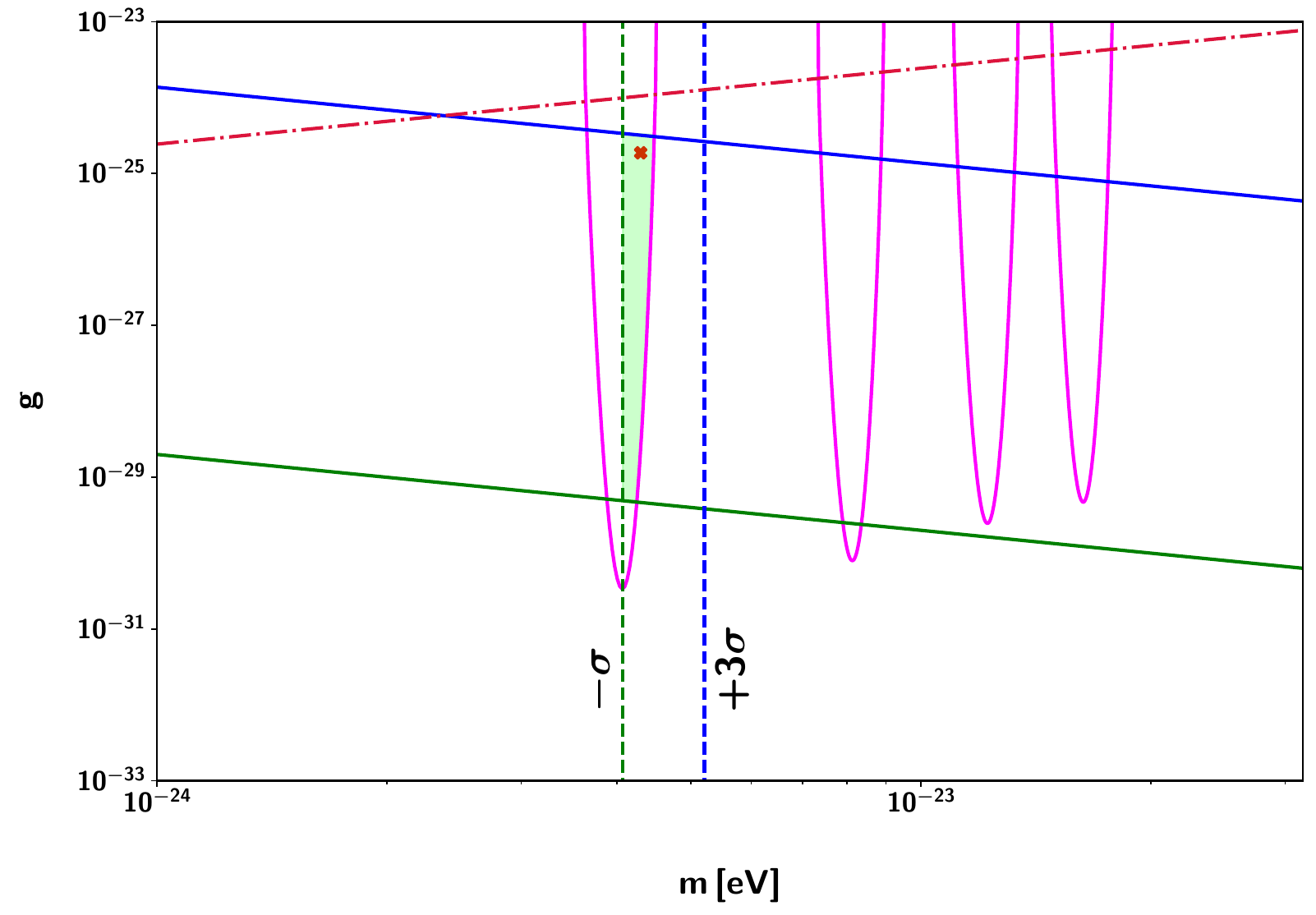}
			\caption{\textit{The allowed parameter space in the $g$ vs $m$ plane based on constraints from the amplitude as well as the shape of the reported signal~\cite{NANOGrav:2023gor}. The vertical dashed lines highlight the constraints on the dark matter mass from the amplitude of the signal. On top of this we project the 3$\sigma$ boundary (represented by the pink contours) obtained from the $\chi^{2}$ test on the excess time delay as depicted in Fig.~1(a) of ref.~\cite{NANOGrav:2023gor}. The lime shaded area highlights the portion of the allowed parameter space that lies within 3$\sigma$ CL.}}
			\label{fig:parameter_g_m_chi}
		\end{figure}

	We also obtain constraints on the mass of the vector dark matter by calibrating the amplitude, corresponding to the combined effect of the Shapiro time delay and the oscillations of the pulsars in the VDM background, against the amplitude of the SGWB signal reported by NANOGrav, i.e.,
		\begin{eqnarray}\label{eq:amplitude-constraint}
			\left(1+\alpha\right)\, \Phi^{\rm GP}_{\rm VDM} = \Phi_{\rm GW}\left(f_i\right).
		\end{eqnarray}
		Using the expressions for $\Phi_{\rm GW}\left(f_i\right)$, $\Phi^{\rm GP}_{\rm VDM}$ given in Eqs.~\eqref{eq:amplitude-SGW} and ~\eqref{Phi-VDM-GP} respectively, we substitute the following values,
		\begin{eqnarray}
			A_{\rm GW} &=& 6.4\times10^{-15}, \qquad \gamma = 3.2, \nonumber\\
			f_{\rm ref} &=& 1\, {\rm yr}^{-1}, \,\, \quad\qquad T_{\rm obs} =\, 16.03 {\rm yrs}. 
		\end{eqnarray}
		Then, setting $f_i = \left(m/\pi\right)$ eV and solving for $m$ in Eq.~\eqref{eq:amplitude-constraint} gives us constant values of $m$ corresponding to a fixed choice of $\alpha$. These constraints are highlighted using the vertical dashed lines in Fig.~\ref{fig:g-vs-m}. The parallelogram formed by the intersection of the upper and lower bounds obtained from the correlation as well as the amplitude defines the $1\sigma$ allowed region in the $g-m$ plane. We note that this narrow range of $(g,m)$ values falls neatly within the broader experimentally allowed region based on the measurements of the time periods of binary pulsars \cite{KumarPoddar:2019ceq}. The boundary separating the allowed and disallowed regions have been demarcated by the crimson dashed-dotted line in Fig.~\ref{fig:g-vs-m}, and the area below the line corresponds to the permitted parameter space. The red cross mark highlights the best-fit point to the NANOGrav data. The frequency associated with the corresponding dark matter mass is around $2\, \rm {nHz}$, which coincides with the peak frequency in the frequency band observed by NANOGrav~\cite{NANOGrav:2023gor}. We have performed a
		direct search in PTA data for both an SGWB signal and a signal from ULDM. We observed that the allowed parameter space in the $g$ - $m$ plane of ULDM lies within
		the 3$\sigma$ confidence level. The pink contours as depicted in Fig.~\ref{fig:parameter_g_m_chi} represent the contours correspond to the 3$\sigma$ confidence level.
		
		\section{Conclusions}\label{sec:conclusions} 
		In this paper, we proposed a non-gravitational explanation of NANOGrav pulsar timing observations~\cite{NANOGrav:2023gor,NANOGrav:2023hvm}. We proposed that $L_\mu-L_\tau$ gauge bosons as the ultralight dark matter, which can have two effects on pulsar timings observed on Earth. Firstly, pulsars carry about $\sim 10^{55}$ muons, therefore they will oscillate with a frequency $f=(m/\pi)$ due to the $L_\mu-L_\tau$ `electric field' of the dark matter. This will give rise to an angular correlation between pulsar residuals proportional to $\frac{1}{2} \cos \theta_{ab}$ as compared to pulsar oscillations due to stochastic gravitational wave background which give the well-known Hellings-Downs angular correlations~\cite{HD}.  The second effect of ultra-light vector dark matter is to cause a gravitational time delay, which also contributes to the angular correlation of pulsar residuals ~\cite{Nomura:2019cvc,Omiya:2023bio}. We found that the combination of these two effects of the $L_\mu-L_\tau$ dark matter gives a better fit to the angular correlation observations of NANOGrav \cite{NANOGrav:2023gor} than the standard SGWB explanation or the SGWB combined with time delay explanations. The mass and couplings of the $L_\mu-L_\tau$ gauge bosons, which give the best fit to the NANOGrav are not ruled out by other astrophysical constraints~\cite{KumarPoddar:2019ceq}. Our analysis shows that in addition to the SGWB signal, there may potentially be excess timing residuals attributable to the $L_{\mu} - L_{\tau}$ ULDM. Future observations of the angular correlation of pulsar timings may be able to enhance the statistical significance of the $L_\mu-L_\tau$ dark matter explanation or rule it out.  
		
		\acknowledgments
		S.P. acknowledges the support by the MHRD, Government of India, under the Prime Minister's Research Fellows (PMRF) Scheme 2020, and by the Spanish Ministerio de Ciencia, Innovación y Universidades through grants PID2020-114473GBI00 and CNS2022-135595. A.H. would like to thank the MHRD, Government of India for the research fellowship. This research of D.C. is
		supported by an initiation grant IITK/PHY/2019413 at IIT Kanpur and by a DST-SERB
		grant SERB/CRG/2021/007579.
		
		\appendix
		\section{$L_\mu-L_\tau$ charge of pulsars}\label{app:muon-charge}
		
		In the neutron stars, electrons are degenerate and there is a beta-equilibrium in the processes $n \rightarrow p+e^-+\bar \nu_e$, $n \rightarrow p+\mu^-+\bar \nu_\mu$ and $\mu^-\rightarrow e^-+\bar\nu_e+\nu_\mu$. The charge neutrality condition constrains the number densities of $p,\, e^-$ and $\mu^-$ to obey the relation $n_p=n_e+n_\mu$. 
		In neutron stars electrons are degenerate and have a Fermi momentum $k_f=(3\pi^2 n_e)^{1/3}$ greater than the mass of the muon $m_\mu\sim 106\,{\rm MeV}$. There is a beta blocking of the outgoing electrons in the muon decay and muons are stable inside neutron stars. The equilibrium of the beta-decay processes in the neutron stars gives the relations between the chemical potential of the different species, i.e., $\mu_n=\mu_p+\mu_e=\mu_p+ \mu_\mu$ which implies that $\mu_e=\mu_\mu$. The chemical potential of the degenerate electrons is the Fermi-energy given by
		\begin{eqnarray}
			\mu_e={\epsilon_e}_f= {k_e}_f =\left(3\pi^2 n_e\right)^{1/3}\,.
			\label{mue}
		\end{eqnarray}
		The muon-chemical potential written in terms of muon number density is
		\begin{eqnarray}
			\mu_\mu = {\epsilon_\mu}_f &=& \left(m_\mu^2 +{{k_\mu}_f}^2\right)^{1/2} \nonumber\\
			&=&\left(m_\mu^2 +\left(3\pi^2 n_\mu\right)^{2/3} \right)^{1/2}\,.
			\label{mumu}
		\end{eqnarray}
		Equating Eqs.~\eqref{mue} and \eqref{mumu} we can solve for $\nu_\mu$ in terms of $\nu_e$ to obtain
		\begin{eqnarray}
			n_\mu=\cfrac{1}{3\pi^2}\left[\left(3\pi^2 n_e\right)^{2/3}-m_\mu^2\right]^{3/2}
			\label{nmu}.
		\end{eqnarray}
		Model calculations give the electron fraction $Y_e=n_e /n= 0.052$ and the nucleon density in neutron stars $n=0.238 {\,\rm fm^{-3}}$ \cite{Pearson:2018tkr}. This gives the number density of electrons in neutron stars $n_e= Y_e\, n= 0.01237\, {\rm fm^{-3}}= 1.237\times 10^{37} \,{\rm cm^{-3}}$. Using this in Eq.~\eqref{nmu}, we obtain the number density of muons in neutron stars to be $n_\mu=3.63 \times 10^{36} {\,\rm cm^{-3}}$ which implies that a typical neutron star with of $R=10 {\,\rm km}$ radius will carry a total muon number 
		\begin{eqnarray}
			N_\mu=n_\mu \left(\cfrac {4\pi}{3} R^3\right)= 1.55 \times 10^{55}\,.
		\end{eqnarray}
		The $L_\mu-L_\tau$ charge of a neutron star is then the coupling constant times the number of muons, $Q=g N_\mu$.
		
		\section{Angular integration over polarizations}\label{app:angular-integ}
		
		The angular integration in Eq.~\ref{eq:integ-angular} can be performed by choosing an arbitrary $\hat{n}$ = $(\sin\theta \, \cos\phi, \sin\theta \, \sin\phi,  \cos\theta)$ and $\epsilon_{1,2}(\vec n)$ as follows:
		
		\begin{eqnarray}
			\epsilon_1(\vec n) = \begin{pmatrix}
				-\sin\phi \\
				\cos\phi \\
				0
			\end{pmatrix}, \quad
			\epsilon_2(\vec n) = \begin{pmatrix}
				\cos\theta \, \cos\phi \\
				\cos\theta \, \sin\phi \\
				-\sin\theta
			\end{pmatrix}.
		\end{eqnarray}
		
		This choice of $\epsilon_{1,2}$ ensures that
		
		\begin{eqnarray}
			\vec{n}\cdot\epsilon_1(\vec n) = \vec{n}\cdot\epsilon_2(\vec n) = \epsilon_1(\vec n)\cdot\epsilon_2(\vec n) = 0.
		\end{eqnarray}
		
		For the pulsars we can choose one of them to be oriented along the $z$-axis, $\vec{n}_a = (0,0,1)$, and the other along the direction $\vec{n}_b = (\sin\theta_{ab}\,\cos\phi_{ab}, \, \sin\theta_{ab}\,\sin\phi_{ab},\, \cos\theta_{ab})$. With these choices, the integrand in Eq.~(8) of the main article becomes
		
		\begin{eqnarray}
			&& \sum_{A=1,2}\, \epsilon_A^i (\vec n)\, \epsilon_A^j(\vec n)\, n^i_a \, n^j_b = \sin^2\theta\,\cos\theta_{ab}
			- \sin\theta\,\cos\theta\,\sin\theta_{ab}\,\cos(\phi - \phi_{ab}).
		\end{eqnarray}
		Integrating first with respect to $\theta$  within the limits $\theta \in [0,\pi]$ gives the result $\cfrac{4}{3}\,\cos\theta_{ab}$. An overall factor of $2\pi$ results from the integration with respect to $\phi$. Thus, we arrive at
		\begin{eqnarray}
			\cfrac{1}{4 \pi} \int \sin\theta\,d\theta\,d\phi\, \sum_A\, \epsilon_A^i (\vec n)\, \epsilon_A^j(\vec n)\, n^i_a \, n^j_b = \cfrac{2}{3} \, \cos \theta_{ab}.\nonumber
		\end{eqnarray}

\end{document}